\documentclass[prl,aps,amsmath,showpacs,reprint,superscriptaddress,floatfix]{revtex4-1}

\usepackage[dvips]{graphicx}
\usepackage[dvips]{color}
\usepackage{amsfonts}
\newcommand{\bra}[1]{\left<{#1}\right|}
\newcommand{\ket}[1]{\left|{#1}\right>}

\newcommand{\HC}[0]{\text{H.C.}}

\begin{document}

\title{On-demand production of entangled photon pairs via quantum-dot biexciton control}

\author{Guy Bensky}
\affiliation{Department of Chemical Physics, Weizmann Institute of Science, Rehovot 76100, Israel}
\author{Selvakumar V. Nair}
\affiliation{Centre for Advanced Nanotechnology, University of Toronto, 170 College St., Toronto M5S 3E3, Canada}
\author{Harry E. Ruda}
\affiliation{Centre for Advanced Nanotechnology, University of Toronto, 170 College St., Toronto M5S 3E3, Canada}
\author{Shubhrangshu Dasgupta}
\affiliation{Department of Physics, Indian Institute of Technology Ropar, Rupnagar, Punjab 140 001, India, and Chemical Physics Theory Group, University of Toronto, Toronto, Canada M5S 3H6}
\author{Gershon Kurizki}
\affiliation{Department of Chemical Physics, Weizmann Institute of Science, Rehovot 76100, Israel}
\author{Paul Brumer}
\affiliation{Chemical Physics Theory Group, University of Toronto, Toronto, Canada M5S 3H6}

\date{\today}

\begin{abstract}

A pulsed laser scenario, designed to prepare an appropriately chosen
self-assembled quantum dot with 100\% population in a biexciton state,
is proposed. The ensuing radiative emission would provide a near perfect
(99.5\%) high rate, on-demand, source of entangled photons. 


\end{abstract}

\maketitle

\paragraph{Introduction --}
Entangled photon generation is of considerable significance in quantum
cryptography where the appearance of unentangled photons in a signal
indicates possible eavesdropping. In such cases extremely high-quality
entangled photon generation (essentially $100\%$ of the emitted
photons being entangled) is required \cite{GisinRMP02}. Other uses of
entangled photons would require less entangled-photon purity (e.g.
$99\%$) and include a number of applications such as two-photon
absorption \cite{DayanPRL05}, microscopy \cite{TeichPAT98},
lithography \cite{SalehPRL05,*BentleyOPTEX04,*FarrerJACS06} and
two-photon coherent tomography \cite{NasrOPTEX04}.  Several of these
applications, such as entangled-photon imaging
\cite{BotoPRL00,*DAngeloPRL01} and entangled photon lithography \cite{PittmanPRA95,*AbouraddyPRL04}, have not been successful in setups where the
entangled photons are generated by standard parametric down conversion,
since  these applications require a high rate of photon-pair
generation. Parametric down conversion
is very ineffective in this respect: photon pairs are generated in
this technique with very small probability.  Hence, there is
considerable interest in developing new, efficient sources of
entangled photons that can prove useful for assorted technical and fundamental applications.

In this paper we propose a nearly-deterministic method for producing
entangled photons at a high rate. It resolves numerous problems associated
with the recent proposal
\cite{BensonPRL00} and demonstration
\cite{StevensonNAT06,*AkopianPRL06} that the decay of quantum-dot
systems from the biexciton state to the ground state can be used as
a triggered source of entangled photon pairs. The quantum dot decays
spontaneously via two exciton states, representing two decay channels.
If the exciton states cannot be spectrally resolved (i.e., when the
splitting between the exciton states due to anisotropic electron-hole
exchange is much less than the radiative linewidths of these states),
then welcher-weg  information on the two photons
is erased.  Thus, the generated photons are entangled in polarization
basis.

To date, this method has generated photon pairs that are very far
from perfectly correlated, due to background noise: The best fidelity
of producing the entangled pairs has been poor, basically limited to $\sim70\%$ even
after reducing the noise \cite{YoungNJP06}. Noting that one of the
main sources of the background noise in these experiments is the
radiative decay of the exciton states, we here propose a method that
avoids such decay. Specifically, we show that it
is possible to achieve {\em complete} population transfer from the
ground state to the biexciton state. In that case, any two consecutive
photons that are generated from the quantum dots would be entangled.
The first photon is generated when the quantum dot decays from the
biexciton to one of the exciton states and the next photon is
generated from the decay of the same exciton state to the ground
state. We also show that Auger effects, which could add unentangled
photon contamination, can be controlled by suppressing recapture of
excited electrons or holes using an electric field.

\begin{figure}
\includegraphics[width=0.8\linewidth]{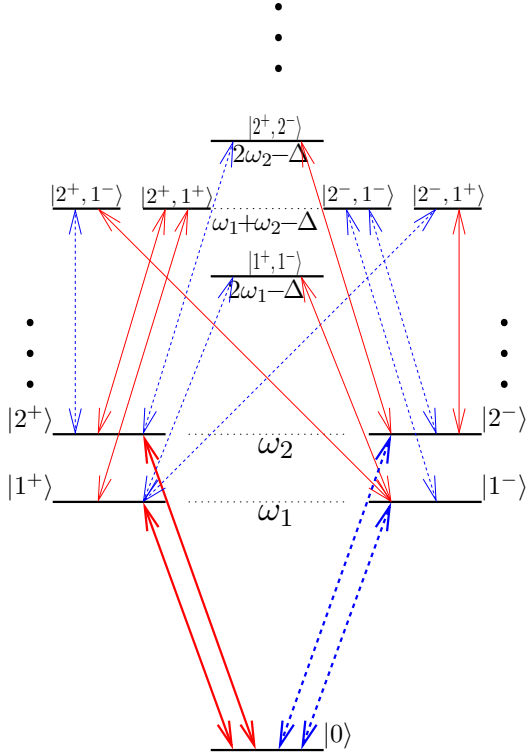}
\caption{
\label{fig-model}
Level structure and radiative transitions of a quantum dot. Only the
first two exciton levels and the corresponding biexciton levels are
relevant in this simulation. The red (solid) arrows denote
transitions of the $\sigma_+$ polarized component, while the blue
(dotted) arrows denote transitions of the $\sigma_-$ polarized
component of the electric field. Thick lines correspond to
ground-to-exciton transitions, while thin lines correspond to
exciton-to-biexciton transitions. $\omega_1$ and $\omega_2$
are the energy levels of the first and second exciton levels
respectively, and $\Delta$ is the biexciton binding energy.}
\end{figure}

\paragraph{Model system --}
\label{sec-model}
Consider a self-assembled quantum dot with multiple exciton states. For each
pair of exciton states there exists a biexciton state, corresponding
to both these states being populated. When subject to a linearly
polarized electromagnetic pulse with field $\vec E(t)$, the ground
state $\ket{0}$ is coupled to the exciton states $\ket{1^+}$,
$\ket{2^+}$, $\dots$ and $\ket{1^-}$, $\ket{2^-}$, $\dots$ by the
$\sigma_+$ and $\sigma_-$ components of the input pulse, respectively.
The same components couple the exciton states to the biexciton states
$\ket{1^+,1^-}$, $\ket{1^+,2^-}$, $\ket{1^+,2^+}$, $\ket{2^+,2^-}$,
etc. with selection rules depicted in Fig.~\ref{fig-model}. The
frequency of the exciton-to-biexciton transition is offset by the
binding energy $\Delta$ from the corresponding ground-to-exciton
transition frequency.  In a typical quantum dot, $\Delta\ll\omega_k$,
where $\omega_k$ is the transition frequency between  states
$\ket{0}\leftrightarrow\ket{k^\pm}$.

It is important to note that nonradiative transitions (notably Auger
recombination) of the quantum dots from the biexciton to the exciton
states serve to reduce the probability of obtaining a pair of photons
that could otherwise be entangled. Thus, it is necessary to identify
systems in which the exciton state left behind after Auger
recombination will not radiate. In a typical self-assembled quantum
dot, the band gap energy is much larger than the confinement barrier
so that the final state of Auger recombination is a highly excited
exciton state with the electron or hole in the continuum of the
barrier material. Recapture of such an excited exciton into the
quantum dot can be suppressed by sweeping the electron-hole pair out
of the system with a dc electric field. It has been demonstrated
  that with such an electric field, even weakly excited electrons can
  be tunneled out of a quantum dot in time scales faster than
  phonon-assisted relaxation \cite{IgnatievPRB01}.  In addition, 
intra-band relaxation of quasi-free carriers in the conduction or
valence band is non-radiative because of energy-momentum conservation
constraints \cite{FriedmanRMP88,*KurizkiPRL93,*ShermanPRL95}.

\paragraph{Fast and efficient population transfer --}
\label{sec-method}
To reduce the noise during generation of the entangled photon pairs,
we propose a method of complete efficient and rapid population
transfer from the ground state $\ket{0}$ to the lowest biexciton state
$\ket{1^+,1^-}$, and  computationally demonstrate its efficiency. The exciton states
$\ket{1^+}$ and $\ket{1^-}$ are coupled to the ground state $\ket{0}$
via the $\sigma_+$ and $\sigma_-$ polarized components of the input pulse,
respectively, and from there to the biexciton state $\ket{1^+,1^-}$ via
the $\sigma_-$ and $\sigma_+$ polarization components, respectively.

Since we want to minimize the time spent in the single exciton states
$\ket{1^+}$ and $\ket{1^-}$ so as to reduce  non-entangled photon emission, the total duration
$T_\text{tot}$ of the pulse must be much shorter than the radiative
decay time $T_\text{tot}\ll\gamma^{-1}\sim1\mathrm{ns}$, where
$\gamma$ is the radiative decay rate. For such short
times, the time energy uncertainty prevents us from distinguishing
between the first exciton states $\ket{1^\pm}$ and the other, higher
states $\ket{2^\pm}\dots\ket{n^\pm}$, where $\omega_n-\omega_1\sim
T_\text{tot}^{-1}$. Thus any computational demonstration of the
photon-pair generation must include these higher exciton levels and
their biexciton counterparts.

The Hamiltonian in the dipole approximation can be written as
\begin{equation}
  \begin{split}
    H= & \sum_{k,s}\hbar\omega_k \ket{k^s}\bra{k^s}
        + \sum_{k}(2\hbar\omega_k-\Delta) \ket{k^+,k^-}\bra{k^+,k^-}\\
       & + \sum_{k>j,s,s^\prime}(\hbar\omega_j+\hbar\omega_k-\Delta)
         \ket{k^s,j^{s^\prime}}\bra{k^s,j^{s^\prime}}\\
       & -\sum_{k,s} d_k E^s(t)\ket{k^s}\bra{0}+ \HC\\
       & -\sum_{k,s}d_{kk} E^s(t)\ket{k^+,k^-}\bra{k^{-s}}+ \HC\\
       & -\sum_{k \neq j, s s^\prime}d_{jk}
          E^s(t)\ket{k^s,j^{s^\prime}}\bra{j^{s^\prime}} + \HC\\
  \end{split}
\end{equation}
where $s$, $s^\prime$ are $+$ or $-$ ($-s$ denotes the sign
  opposite to $s$), $E^+$ and $E^-$ are the $\sigma_+$ and $\sigma_-$
polarized components of the electric field $\vec E(t)$ of the input
pulse, $d_k$ is the transition dipole moment between the ground level
$\ket{0}$ and the exciton levels $\ket{k^+}$ and $\ket{k^-}$, and
$d_{jk}$ is the transition dipole moment between the exciton states
$\ket{j^\pm}$ and the biexciton state $\ket{j^\pm,k^s}$. We note that
$\ket{j^s,k^{s^\prime}}$ and $\ket{k^{s^\prime},j^{s}}$ represent
identical states.

The calculations rely on the following conditions,
 appropriate for a typical lens-shaped InGaAs/GaAs
self-assembled quantum dot: (i) We assume that
the quantum dot is initially prepared in the ground state $\ket0$ 
with a typical transition frequency of
$\hbar\omega_1=1.3$ eV
\cite{BayerPRB02,DekelPRB00,*BesterPRB03}. (ii) We limit ourselves to
one additional exciton level $\ket{2^\pm}$, that is $\sim$40 meV
above the exciton ground state \cite{DekelPRB00,*BesterPRB03}.
The biexciton binding energy is taken as   a typical value of
$\Delta=4$ meV for all biexciton states.  (iii) The
ground-to-exciton transitions (with dipole moments $d_k$) in most
quantum dots are an s-s transition to the first exciton state, and a
p-p transition to the second exciton state.  Hence, to first
approximation, the transition dipole moment $d_2$ to the second
exciton state $\ket{2^\pm}$ is $\sqrt{2}$ times stronger than the
transition dipole moment $d_1$ to the first exciton state
$\ket{1^\pm}$. (iv) Calculations that treat the electron-hole
correlation accurately show that the biexciton-to-exciton radiative
decay is $\sim$1.6 to 1.8 times faster than the radiative decay of the
exciton in the size range of interest \cite{WimmerPRB06}.  When
degeneracy (two channels for emission) is accounted for, this
corresponds to a ratio of $0.8$ to $0.9$ for oscillator strengths. For
the proof-of-concept we chose a constant oscillator strength ratio of
$0.8$ between all the exciton-to-biexciton transitions and the
corresponding ground-to-exciton transitions (i.e., $d_{jk} = \sqrt{0.8}d_k$).

The field $\vec E(t)$ is chosen  to be linearly polarized, such that
$E^+(t)=E^-(t)$ and two pulses $E(t)=E_1(t)+E_2(t)$ are applied: a pulse
$E_1(t)$ with frequency $\omega_1$ resonant with the
$\ket{0}\leftrightarrow\ket{1^\pm}$ transition, and another pulse
$E_2(t)$ with frequency $\omega_1-\Delta$ resonant with the
$\ket{1^\pm}\leftrightarrow\ket{1^+,1^-}$ transition. Each pulse has a
Gaussian profile. The duration of the pulses $\tau$ should be long
enough to keep the second exciton state $\ket{2^\pm}$ out of the
time-energy uncertainty regime:
$\tau\gtrsim2\pi\hbar/40~\mathrm{meV}\approx100~\mathrm{fs}$.
Additionally, because the ground-to-exciton transition and the
exciton-to-biexciton transition have different dipole amplitudes, it
is important to access each one separately with a different amplitude
pulse. Hence, to keep them out of the time-energy uncertainty regime,
we must also demand that the pulse times be larger than 
$\tau\gtrsim2\pi\hbar/4~\mathrm{meV}\approx1000~\mathrm{fs}$.

On the other hand,  the transfer needs to be completed well within the
radiative decay time of the exciton levels
$\gamma^{-1}\approx1~\mathrm{ns}$, in order to prevent  decay from occurring
{\em during} the transfer process while the population is in a single
exciton state, as this will generate  unentangled photons.

\paragraph{Analysis --}
\label{sec-analysis}
The scenario described above is here analyzed  for two distinct pulse schemes.

\begin{figure}
\includegraphics[width=\linewidth]{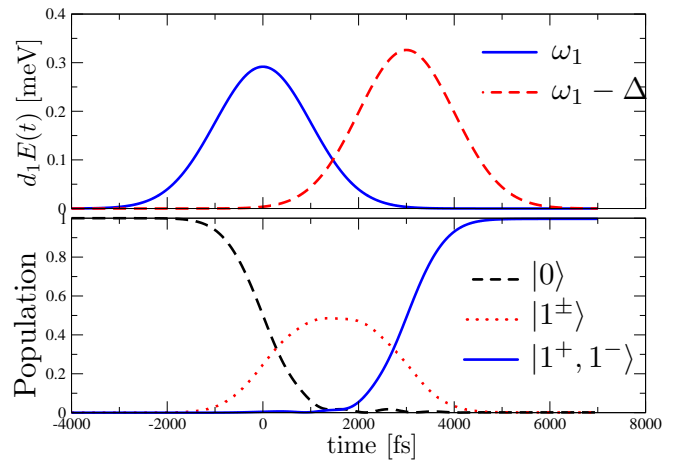}
\caption{
\label{fig-sim-sequential}
Simulation results for a scheme with two sequential Gaussian Rabi
pulses of width $1000\mathrm{fs}$. The first pulse of frequency
$\omega_1$, in resonance with the ``ground-to-exciton''
$\ket{0}\leftrightarrow\ket{1^\pm}$ transition, transfers the
population from the ground to the single exciton states. The following
pulse of frequency $\omega_1-\Delta$, in resonance with the
``exciton-to-biexciton'' $\ket{1^\pm}\leftrightarrow\ket{1^+,1^-}$
transition, transfers the population from the single exciton states to
the biexciton state. The result of this scheme leaves $99.7\%$ of the
population in the biexciton state and $\sim0.1\%$ of the population in
the ground state, resulting in at most only $\sim0.2\%$ non-entangled
(``bad'') photons generated. During the transfer the population spends
a total of $3\mathrm{ps}$ in the single exciton states.}
\end{figure}

{\em (a) Sequential-pulse scheme}, where the ``ground-to-exciton''
pulse $E_1(t)$ {\em precedes} the ``exciton-to-biexciton'' pulse
$E_2(t)$. In this scheme,  all the population is first transferred 
from the ground state $\ket{0}$ to the single exciton states
$\ket{1^+}$ and $\ket{1^-}$.  The population is then excited 
from these single exciton states to the biexciton state $\ket{1^+,1^-}$.
The advantage of such a scheme is that the two pulses $E_1(t)$ and
$E_2(t)$ hardly interfere \cite{KurizkiPRB89} insofar as they have little
overlap in time.  However, the disadvantage  is that
 the system is in a single exciton state between pulses, where it can
decay by emitting a single, non-entangled photon, thereby reducing the
fidelity of the generated entangled photons. Nevertheless we
  find that using 1000 fs pulses
  [Fig.~\ref{fig-sim-sequential}], $99.7\%$ of the population is in
the biexciton state $\ket{1^+,1^-}$,  $\sim0.1\%$ in the ground state
$\ket{0}$.  Only $\sim0.2\%$ in one of the other
states, which can result in non-entangled photons.  The population
spends a total of 1.5 ps in each single exciton state
$\ket{1^\pm}$, resulting in
$\sim 3~ \mathrm{ps}\times\gamma\approx0.3\%$  additional 
non-entangled photons.

\begin{figure}
\includegraphics[width=\linewidth]{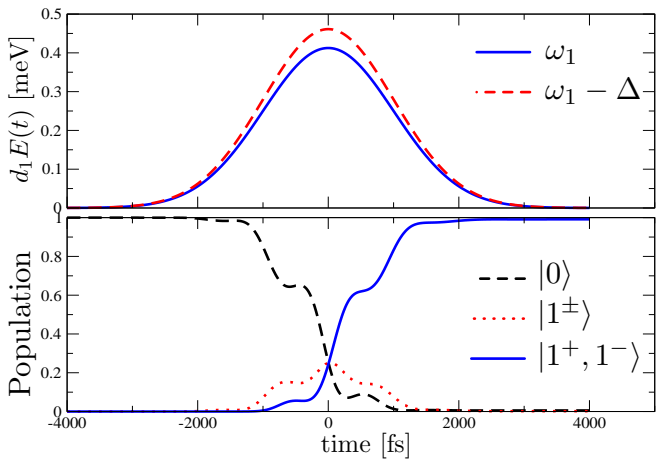}
\caption{
\label{fig-sim-concurrent}
Simulation results for a scheme with two concurrent Gaussian Rabi
pulses of width $1000\mathrm{fs}$. One pulse is of frequency
$\omega_1$, in resonance with the ``ground-to-exciton''
$\ket{0}\leftrightarrow\ket{1^\pm}$ transition, and the other pulse is
of frequency $\omega_1-\Delta$, in resonance with the
``exciton-to-biexciton'' $\ket{1^\pm}\leftrightarrow\ket{1^+,1^-}$
transition. Both together transfer the population from the ground
state, through the single exciton states, to the biexciton state. The
result of this scheme leaves $99.1\%$ of the population in the
biexciton state and $\sim0.5\%$ of the population in the ground state,
resulting in at most only $\sim0.4\%$ non-entangled (``bad'') photons
generated. During the transfer the population spends a total of
$0.7\mathrm{ps}$ in the single exciton states. Note the oscillatory nature of the population transfer, compared with the
sequential transfer in Fig.~\ref{fig-sim-sequential}, a feature that is due to
the interference  inherent in such concurrent pulses.
}
\end{figure}

{\em (b) Concurrent pulses scheme}. Here the ``ground-to-exciton''
pulse $E_1(t)$ and ``exciton-to-biexciton'' pulse $E_2(t)$
overlap, transferring the population simultaneously from ground
$\ket{0}$ to the exciton states $\ket{1^+}$ and $\ket{1^-}$, and from
the exciton states to the biexciton state $\ket{1^+,1^-}$. The advantage
of such a scheme is that the population remains only a minimal amount
of time in the single exciton state, thus minimizing the probability
of non-entangled photon decay. The disadvantage is that
%
due to interference effects between the pulses,
the populations oscillate in time  more dramatically than in
the sequential-pulse scheme, resulting in
%
the (slightly) lower fidelity
[Fig.~\ref{fig-sim-concurrent}]. The best result  found for the
population using 1000 fs pulses is $99.1\%$ in the biexciton
state $\ket{1^+,1^-}$ and $\sim0.5\%$ in the ground state $\ket{0}$
which leaves $\sim0.4\%$ in one of the other states, which can
  result in non-entangled photons.  The population spends a total of
0.35 ps in each single exciton state $\ket{1^\pm}$,
resulting in 0.7 ps $\times\gamma\approx0.07\%$ additional
non-entangled photons.

Hence, rather unexpectedly, the sequential-pulses scheme performs better, despite the
longer time spent in the single exciton state. However, for systems
with a larger radiative decay rate $\gamma$, e.g. a system with
$\gamma^{-1}\approx0.1$ ns, the concurrent-pulses scheme might
be better since the added decay probability from a single-exciton
  state in the sequential-pulses scheme will be larger, negating the advantage of this scheme over the concurrent-pulses scheme.

\paragraph{Summary --}
\label{sec-conclusion}
In conclusion, we propose a highly effective population transfer technique to avoid
radiative  decay of exciton states in a quantum dot and thus
dramatically increase the fidelity of generating entangled photon
pairs by spontaneous decay of the biexciton state. When the quantum
dot is prepared by appropriate pulses in the biexciton states, with no
initial population in the exciton states, contamination due to exciton decay
can be avoided. Any two consecutive photons that would be emitted due
to cascade decay from the biexciton state would then be entangled in the 
($\sigma_\pm$) polarization basis. We showed, using two laser pulses, how to
prepare the quantum dot in the biexciton state and that $99.8\%$ entangled photon
generation is achievable using realistic system parameters. Higher
fidelity of transfer is more challenging but not impossible, and is best carried 
out using optimal control techniques in conjunction with the specific 
experimental arrangement that is adopted.

Finally, we noted that in a quantum dot, the probability of nonradiative decay
from biexciton to exciton states due to Auger effects is not
completely eliminated, but that this can be overcome by choosing dots
where the resulting high-energy exciton is above the confinement
energy, whereby it can be removed from the system, as discussed above.

The realistic requirements  of the proposed scheme make it a viable
  method for nearly-deterministic, high-rate, on-demand production of entangled
  photon pairs, opening the possibility of  a variety of quantum
  applications
  \cite{DayanPRL05,TeichPAT98,SalehPRL05,*BentleyOPTEX04,*FarrerJACS06,NasrOPTEX04,BotoPRL00,*DAngeloPRL01,PittmanPRA95,*AbouraddyPRL04}.

\paragraph{Acknowledgement}  PB and SD 
 thank Professor G. D. Scholes for various discussions during the early
stages of this work.  Support from the Natural Sciences and Engineering Research Council of Canada is gratefully acknowledged.

\bibliography{Bibliography}

\end{document}